\documentclass[preprint,11pt]{JHEP3} 

\JHEPspecialurl{http://jhep.sissa.it/JOURNAL/JHEP3.tar.gz}

\usepackage{epsfig,multicol,amsmath,slashed}
\usepackage[square, comma, sort&compress]{natbib}

\def\beq{\begin{equation}}
\def\eeq{\end{equation}}
\def\bea{\begin{eqnarray}}
\def\eea{\end{eqnarray}}

\def\eq#1{{Eq.~(\ref{#1})}}
\def\fig#1{{Fig.~\ref{#1}}}

\def\Tab#1{{Table~\ref{#1}}}

\parskip=\itemsep               

\newcommand{\nn}{\nonumber}

\newcommand{\h}{\frac{1}{2}}

\newcommand{\al}{\alpha}

\newcommand{\de}{\delta}

\newcommand{\ka}{\kappa}
\newcommand{\la}{\lambda}

\newcommand{\Ga}{\Gamma}

\newcommand{\La}{\Lambda}

\renewcommand{\theequation}{\thesection.\arabic{equation}}
\newcommand{\f}{\frac}

\newcommand{\SPE}{\langle \mid S_{\mbox{\footnotesize{enh}}}^2 \mid\rangle}
\newcommand{\If}{\int^{\infty}_{-\infty}}

\newcommand{\bas}{\bar{\alpha}_s}
\newcommand{\as}{\alpha_s}

\newcommand{\Lb}{\left(}
\newcommand{\Rb}{\right)}

\newcommand{\GV}{\mbox{GeV}/\mbox{c}^2}

\title{\LARGE \bf Pomeron loop summation in perturbative QCD and the survival probability}
\author{\large  J.~Miller\thanks{Email:
jeremymi@post.tau.ac.il; jeremy.miller@ist.utl.pt}\,\, \\
CENTRA, Departamento de F$\acute{i}$sica, Instituto Superior T$\acute{e}$cnico (IST),\\  Av. Rovisco Pais,\\1049-001 Lisboa,\\Portugal}

\abstract{The survival probability for exclusive diffractive Higgs production is calculated. The contribution
of short distance interactions are taken into account, by summing over Pomeron loops in perturbative QCD.
The summation is performed by developing an iterative technique to sum over loop diagrams with higher and higher generations of loops.
The results show that the survival probability depends inversely on energy and is small for the LHC range of energies, and could be even less than 1 \%. 
}

\preprint{ \today}

\begin{document}

\renewcommand{\theequation}{\arabic{equation}}

The most important event anticipated at the LHC is the detection of the Higgs boson, in proton proton collisions. The most desirable result
is the exclusive production of the Higgs with no other particles produced in the scattering. As such there are large rapidity gaps (LRG) between the Higgs and the emerging protons shown
in \fig{flrg1}.  Thanks to the large rapidity gaps, exclusive Higgs production has an excellent experimental signature, and offers the best chance for successfully isolating the Higgs boson. \fig{flrg1} is a double t-channel gluon
exchange leading to a zero net color flow in the t-channel, which cancels the possibility of additional inelastic scattering apart from the Higgs. The evolution of ladder gluons between the two t channel gluons
forms the structure of the BFKL Pomeron in the leading log approximation. 
Unfortunately in high energy scattering, the production of extra unwanted particles from more parton showers, could spoil the large rapidity gaps as shown in \fig{flrg2}, which means detecting the Higgs is problematic.
The survival probability is the probability of having large rapidity gaps between the Higgs and the emerging protons, and as such its value characterizes the probability for exclusive
Higgs production without further unwanted production.

\DOUBLEFIGURE[h]{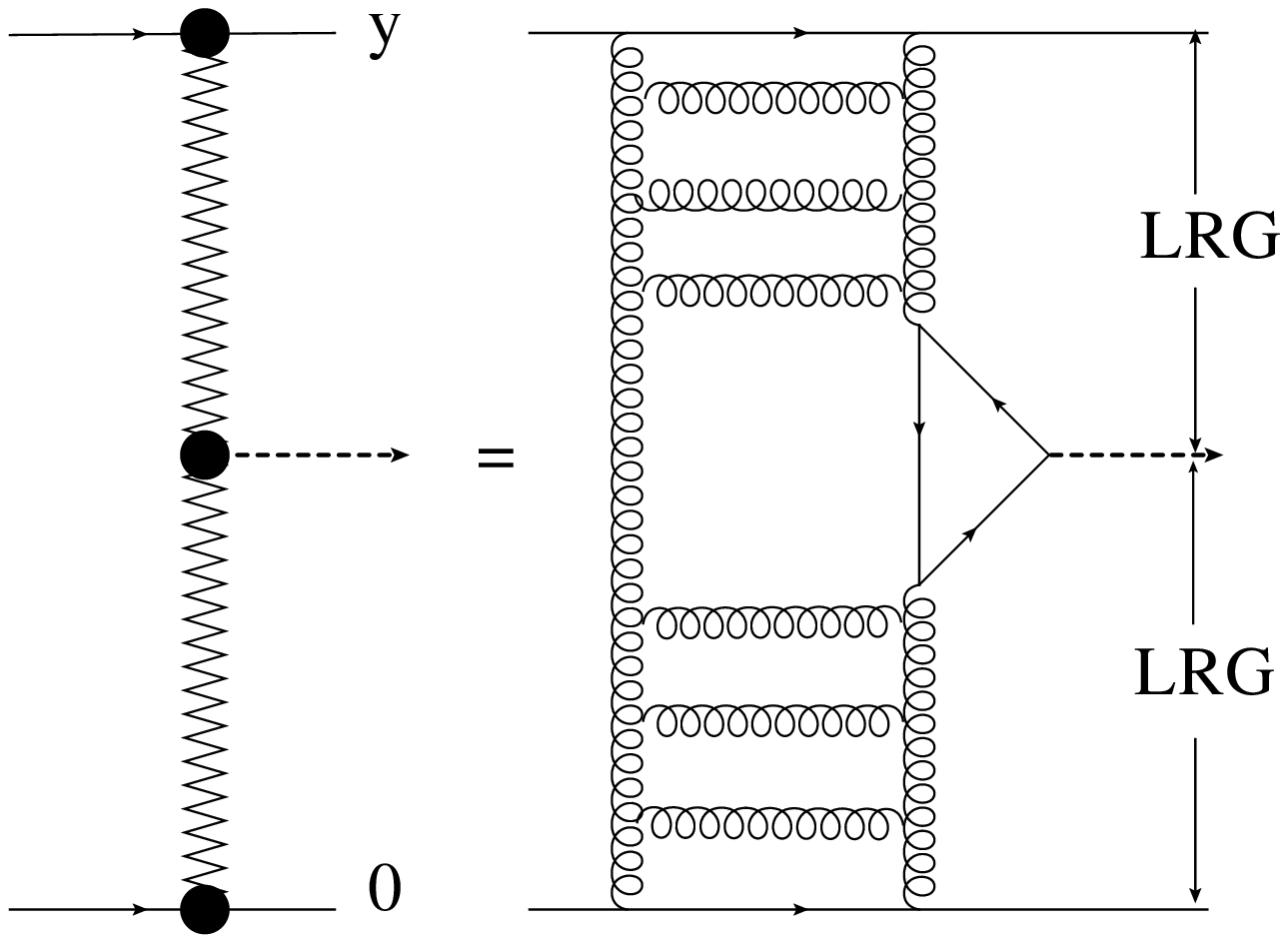,width=60mm,height=45mm}{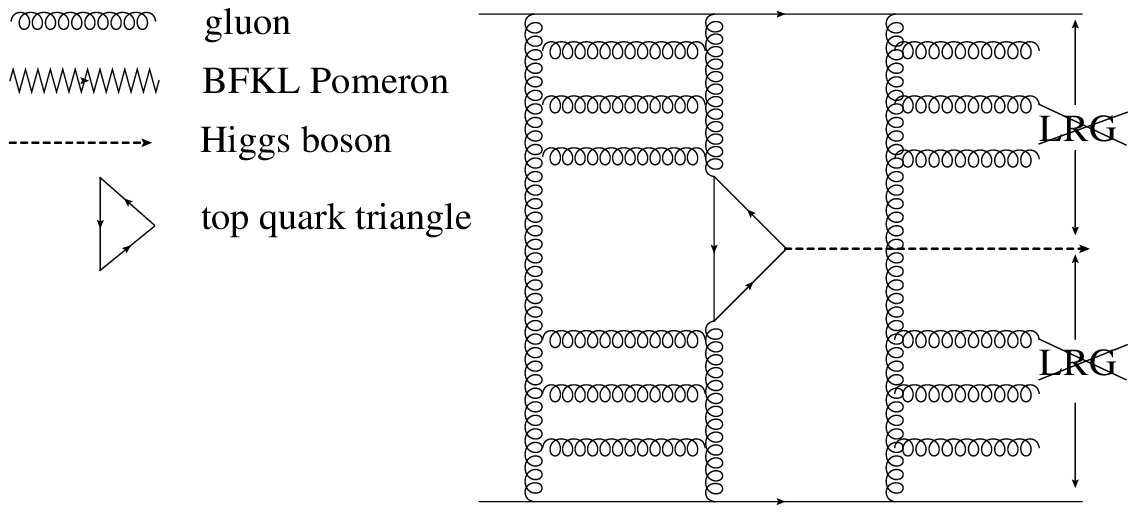,width=110mm,height=45mm}
{Exclusive Higgs production from t-channel BFKL Pomeron exchange, with large rapidity gaps (LRG) between the Higgs and the emerging protons.
\label{flrg1}}{The production of Higgs with extra production which spoils the LRGs, arising from
additional inelastic scattering..\label{flrg2} }

The contribution of the full set of parton showers that destroy the LRGs is required for estimates of the survival probability. Both long and short distance interactions contribute, so a 
reliable estimate is difficult to obtain. The contribution of short distance interactions  can be treated in perturbative QCD, and for
high density protons this type of hard  process can be treated by summing over ladder diagrams of the type shown in \fig{flrg1}. In this approach
\fig{flrg1} is the sum over all diagrams with $n$ rungs of the ladder. The vertical lines of the ladder are themselves a superposition of the sum
over $n$ rung ladder diagrams  and so on. This sum over ladder diagrams leads to the scattering amplitude in the leading log approximation (LLA) which is proportional to 
\cite{Gribov:1983,
Bartels:1975,Lipatov:1989,Cheng:1976,Ross}
$A\Lb s,t\Rb\propto\sum^\infty_{n=0}\Lb\al\Lb t\Rb\ln\Lb s/s_0\Rb\Rb^n/n!\,=s^{\al\Lb t\Rb}$,
where $\al\Lb t\Rb$ is the Regge trajectory.
Hence the sum over ladder diagrams is achieved by replacing
 the pair of interacting vertical gluons in \fig{flrg1} by a ``reggeon"  which behave as $s^{\al\Lb t\Rb}$  for energy $s\gg1$.  According to the optical theorem
the  total cross section  behaves as $\sigma_{tot}\propto s^{\al\Lb 0\Rb-1}$.  Experimentally it is known that the total cross section rises slowly with $s$ which means
 $\al\Lb 0\Rb >1$.  Pomeranchuk
 first commented \cite{Pomeranchuk:1956,Okun:1956} that this behavior is matched by the theoretical prediction that $\al\Lb 0\Rb>1$ when the t-channel exchange carries zero quantum numbers,
 including zero charge and color flow in the t-channel. Such particles with quantum numbers of the vacuum exist in QCD
 for bound gluon states. This trajectory is called the Pomeron named after Pomeranchuk, which is the double t-channel gluon exchange shown in \fig{flrg1}. 
The evolution of the vertical t-channel gluons  to the sum over ladder diagrams is called the BFKL Pomeron which is described by the BFKL equation \cite{Fadin:1976,Balitsky:1978}. \\

The BFKL Pomeron splits and re-merges via the triple Pomeron vertex which was first calculated by Korchemsky in ref. \cite{Korchemsky:1997fy} and by Bialas, Navelet and Peschanski (BNP) in ref. \cite{Bialas:1997ig} .
The large contribution of the triple Pomeron vertex means that Pomeron loop diagrams yield a significant contribution to the scattering amplitude, which is comparable to the basic amplitude of \fig{flrg1} (see for example estimates in
refs. \cite{Miller:2006bi,Levin:2007wc,Miller:2009ca}). Hence it turns out that an accurate estimate for the scattering amplitude should take into account the contribution from Pomeron loop diagrams. To approach this problem
requires developing an approach for summing over Pomeron loop diagrams. Fortunately, we can treat this theoretically in perturbative QCD. The sum over Pomeron loop diagrams provides the complete set of shadowing corrections 
to the basic diagram of \fig{flrg1} arising from hard re-scattering, which spoil the large rapidity gaps. Thus the formalism for summing over Pomeron loops provides the framework for estimating the contribution
of short distance interactions to the survival probability. 

\FIGURE[h]{\begin{minipage}{140mm}
\centerline{\epsfig{file=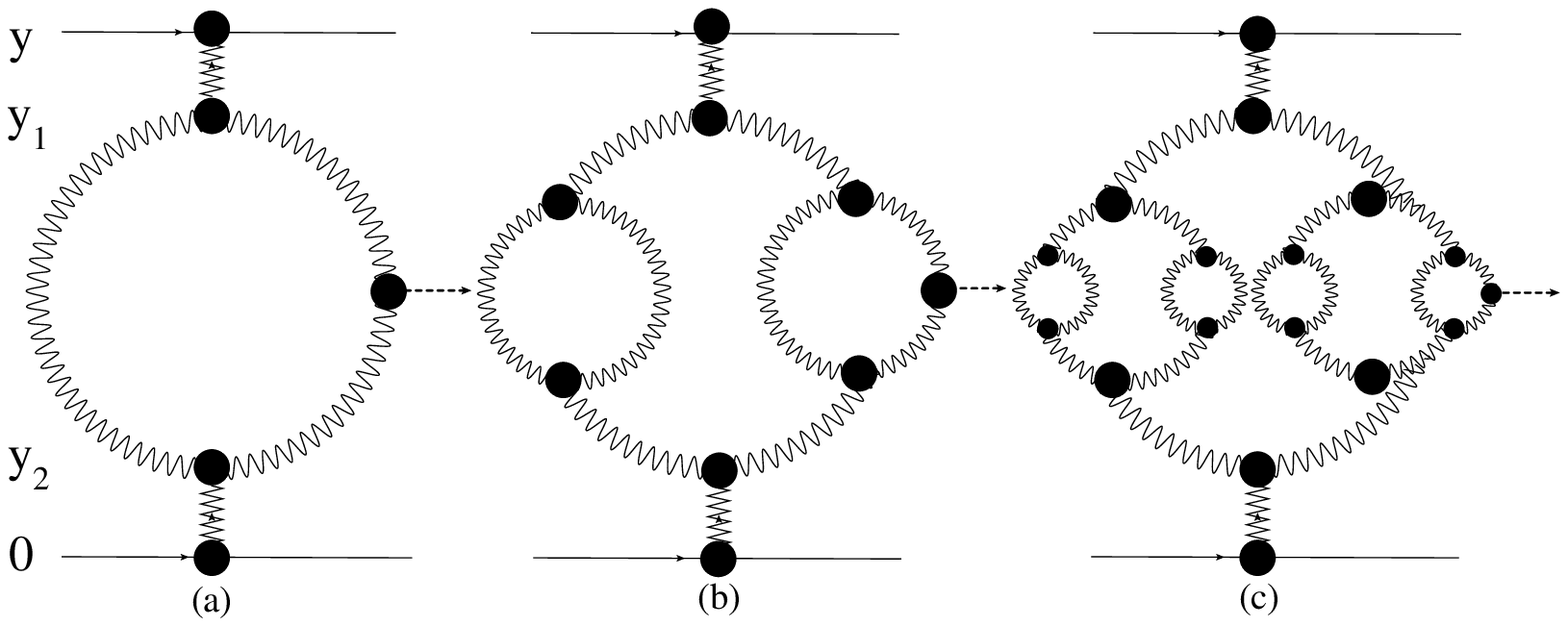,width=140mm}}
\end{minipage}
\caption{The special class of symmetric Pomeron loop diagrams taken into account in the sum over Pomeron loops.} \label{fsopl} } 

In this letter, the special class of Pomeron loops shown in \fig{fsopl} have been taken into account in the summation over Pomeron loops. Diagrams of this type can be calculated 
from the observation that when each branch of the loop in \fig{fsopl} (a) gives birth to a loop, one obtains \fig{fsopl} (b). In this way a second
generation of loops has been introduced, and \fig{fsopl} (b) is called the $N=2$ generation diagram.  \fig{fsopl} (a) with one loop is called the $N=1$ generation diagram. Likewise, a third generation of loops has been introduced
in \fig{fsopl} (c) when each of the 2 simple loops at the center of \fig{fsopl} (b) give birth to two loops in the same way, so  \fig{fsopl} (c) is called the $N=3$ generation diagram. Continuing in this way, the full spectrum of
symmetric Pomeron loop diagrams is generated for all $N$ generation diagrams. In this letter, an iterative algorithm is
described for calculating such diagrams. \\

The solution to the BFKL equation
provides the trajectory $\omega\Lb n,\nu\Rb$ $(n\in \mathbb{Z};\nu\in\mathbb{R})$ for the BFKL Pomeron as \cite{Fadin:1976,Balitsky:1978};

\bea
\omega\Lb n,\nu\Rb=\bas\Lb\psi\Lb 1\Rb-\Re e\,\psi\Lb \f{1+n}{2}+i\nu\Rb\Rb;\hspace{1cm}\bas\equiv\f{\as N_c}{\pi}\label{BFKLeigenvalue}\eea

$n$ represents the energy levels of the BFKL Pomeron, and $\nu$ is a continuous variable which one integrates over when calculating Feynman diagrams. The BFKL eigenfunction falls sharply with increasing $n$ and
is only positive at high energy when $n=0$. Hence throughout this letter which is focussed on high energy scattering, $n=0$ is assumed and the argument $n$ is suppressed.  Hence the BFKL Pomeron trajectory 
which is the sum over ladder diagrams of the type shown in \fig{flrg1} is described by the regge behavior $s^{\omega\Lb \nu\Rb}\equiv e^{\omega\Lb \nu\Rb y}$. The scattering amplitude 
of \fig{flrg1} is well known and has been calculated in refs. \cite{Miller:2006bi,Levin:2007wc,Miller:2009ca,Kozlov:2004sh,Navelet:2002zz,Navelet:1998yv,Navelet:1997xn};

\bea
&&A_{(0)}\Lb y,\de y_H|\mbox{\fig{flrg1}}\Rb\!=\!\f{\as^2}{4}A_H\!\int^\infty_{-\infty}\!\!\!d\nu h\Lb\nu\Rb\la\Lb\nu\Rb e^{\omega\Lb \nu\Rb \Lb y-\de y_H\Rb}E_\nu E^\prime_{-\nu}\label{A0}\\
&&h\Lb\nu\Rb=\f{2}{\pi^4}\nu^2;\hspace{0.5cm}\la\Lb\nu\Rb=\f{1}{16}\f{1}{\Lb 1/4+\nu^2\Rb^2};\hspace{0.5cm}E_\nu=\Lb\f{r_{12}}{r_{10}r_{20}}\Rb^{1/2+i\nu}\Lb\f{r^\ast_{12}}{r^\ast_{10}r^\ast_{20}}\Rb^{1/2-i\nu}\label{hla}\eea

 $h\Lb\nu\Rb$ is the integration measure which preserves conformal invariance \cite{Braun:2009sh,Braun:2005bv}, $\la\Lb\nu\Rb$ is the Pomeron propagator in the conformal basis \cite{Braun:2009sh,Braun:2005bv} and $E_\nu$ is the coupling
 of the BFKL Pomeron to the QCD color dipole \cite{Navelet:1997xn,Braun:2009sh,Braun:2005bv}, in the dipole approach to proton proton scattering. Here $r_{12}=r_1-r_2$ is the transverse size of the dipole and $r_{10}=r_1-r_0$
 where $r_0$ is the center of mass coordinate of the dipole.  $\de y_H=\ln \Lb M_H^2/4 s_0\Rb\,\Lb s_0=1\GV\Rb$ is the rapidity gap occupied by the heavy Higgs boson, and $A_H=\Lb N_c^2-1\Rb\as 2^{1/4} G_F^{1/2}/3\pi$;
(where $G_F$ is the Fermi coupling)  is the
 contribution to the scattering amplitude of the process Pomeron+Pomeron $\to$ Higgs derived in refs. \cite{Miller:2007pc,Ellis:1975ap,Ellis:1976yc,2,Rizzo:1979mf,Dawson:1990zj,22},
  which as shown in \fig{flrg1} proceeds mainly via the intermediate top quark triangle. The observation that the BFKL eigenfunction \eq{BFKLeigenvalue} has a saddle point $\nu=0$ means that one can expand 
  the exponential in \eq{A0} as 
  
  \bea
 \omega\Lb\nu\Rb=\omega\Lb 0\Rb-\h\nu^2\omega^{\prime\prime}\Lb 0\Rb+\mathcal{O}\Lb\nu^2\Rb;\hspace{0.5cm} \omega\Lb 0\Rb=4\bas\ln 2;\hspace{0.5cm}\omega^{\prime\prime}\Lb 0 \Rb=28\bas\zeta\Lb 3\Rb\,.
 \label{BFKLexpansion}\eea
  
 where $\zeta\Lb 3\Rb=1.202$ is the Riemann zeta function. Using this expansion the integration in \eq{A0} is evaluated by the steepest descent method which yields the result \cite{Miller:2009ca};
  
  \bea
A_{(0)}\Lb y,\de y_H|\mbox{\fig{flrg1}}\Rb\!&&=\!\f{\bas^2\Lb 2\pi\Rb^{1/2}\,A_H}{2\pi^2N_c^2}\! \f{e^{\omega\Lb 0\Rb \Lb y-\de y_H\Rb}}{\Lb\omega^{\,\prime\prime}\Lb 0\Rb\Lb y-\de y_H\Rb\Rb^{3/2}}\label{A01}\eea
 
 \eq{A01} is the bare scattering amplitude for the desired process of \fig{flrg1}. 
 The first order shadowing correction is the Pomeron loop shown in \fig{fsopl} (a). Using the same conventions, the scattering amplitude is given by the expression \cite{Miller:2009ca};
 
 \begin{subequations}\label{loop}
 \bea
&&A_{(1)}\Lb y,\de y_H|\mbox{\fig{fsopl} (a)}\Rb=\f{\as^2}{4}A_H\!\If\!\!\!d\nu h\Lb\nu\Rb\la^2\Lb\nu\Rb e^{\omega\Lb\nu\Rb\Lb y-y_{12}\Rb}d_{(1)}\Lb\nu|y,\de y_H\Rb E_\nu E^\prime_{-\nu}\label{loop1}\\
&&d_{(1)}\Lb\nu|y,\de y_H\Rb=\f{1}{16}\If\!\!\! d\nu_1 h\Lb\nu_1\Rb\la\Lb\nu_1\Rb\If\!\!\! d\nu_2 h\Lb\nu_2\Rb\la\Lb\nu_2\Rb|\Ga\Lb\nu|\nu_1,\nu_2\Rb|^2\label{loop2}\\
&&\times\int^y_{\de y_H}\!\!\! dy_1\!\int^{y-\de y_H}_0\!\!\!\!\!\! dy_2 e^{\Lb\omega\Lb\nu_1\Rb+\omega\Lb\nu_2\Rb-\omega\Lb\nu\Rb\Rb y_{12}-\omega\Lb\nu_1\Rb\de y_H}\nn
 \eea
 \end{subequations}
 
 \FIGURE[h]{\begin{minipage}{70mm}
\centerline{\epsfig{file=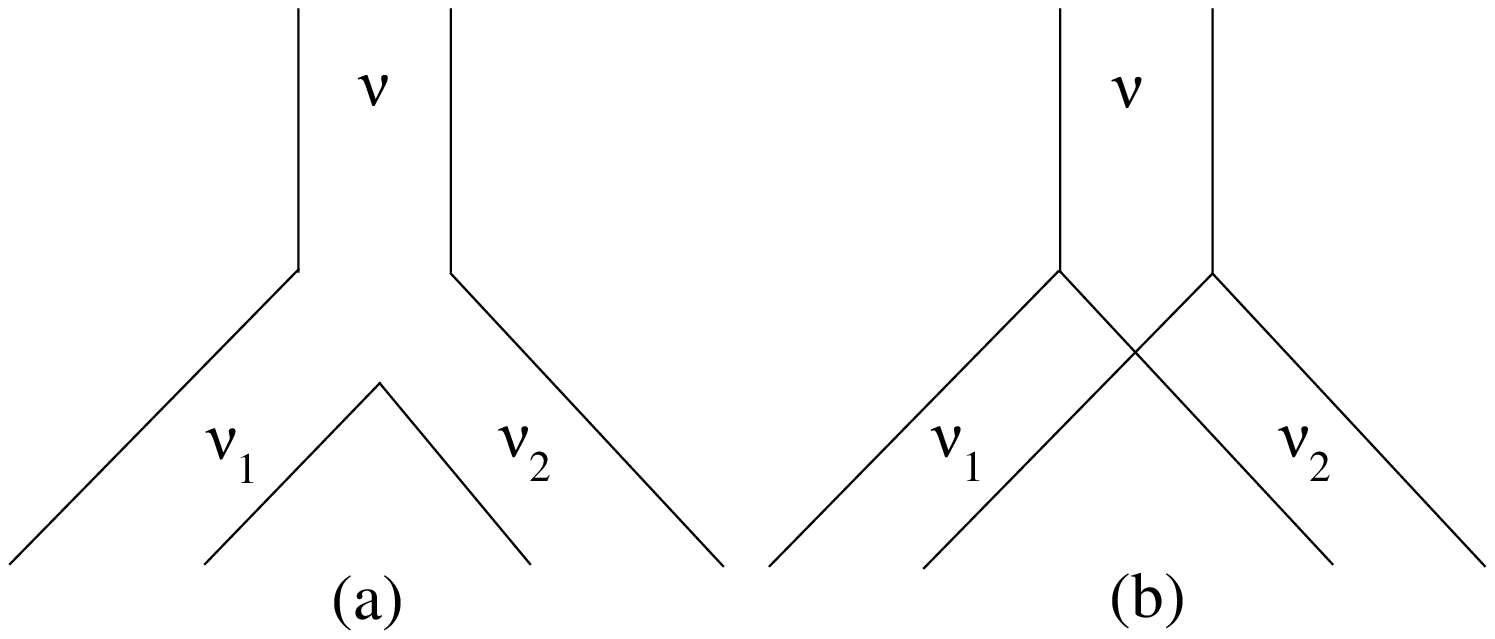,width=70mm}}
\end{minipage}
\caption{The triple Pomeron vertex, where lines indicate ``reggiezed" gluons.  (a) is the planar diagram, and (b) is the non planar diagram. } \label{ftpv} } 

where $y_{12}=y_1-y_2$ is the rapidity gap which the loop fills (see \fig{fsopl} (a)).  $|\Ga\Lb\nu|\nu_1,\nu_2\Rb|^2$ is the triple Pomeron vertex (TPV) 
 for the splitting of a Pomeron with conformal variable $\nu$ into two  Pomerons 
with conformal variables $\nu_1$ and $\nu_2$, shown in \fig{ftpv}.
The explicit expression is very complicated and its full form and the derivation can be found in refs. \cite{Korchemsky:1997fy,Bialas:1997ig}. The TPV is the sum of the planar and non 
planar diagrams shown in \fig{ftpv} (a) and \fig{ftpv} (b), namely;

\bea
\Ga\Lb\nu,\nu_1,\nu_2\Rb&&=16\bas^2 \Lb \f{1}{4}+\nu^2\Rb^2\Lb\underbrace{\Omega\Lb\nu,\nu_1,\nu_2\Rb}+\f{2\pi}{N_c^2}\underbrace{\La\Lb\nu,\nu_1,\nu_2\Rb\Lb\chi\Lb\nu\Rb-\chi\Lb\nu_1\Rb-\chi\Lb\nu_2\Rb\Rb}\Rb;
\label{tpv}\\
&&\hspace{3.7cm}\mbox{\footnotesize{\fig{ftpv} (a)}}\hspace{3.7cm}\mbox{\footnotesize{\fig{ftpv} (b)}}\nn\\
\chi\Lb\nu\Rb&&=\Re e\Lb\psi\Lb 1\Rb-\psi\Lb\h+i\nu\Rb\Rb\nn\eea

In ref. \cite{Miller:2009ca} it was shown that
the two contributions to \eq{loop} stem from the regions I; $\left\{\nu,i\nu_1,i\nu_2\right\}=\left\{ 0,1/2,1/2\right\}$ and region II; $\left\{i\nu,\nu_1,\nu_2\right\}=\left\{ 1/2,0,0\right\}$, for which the TPV
takes the following asymptotic form;

\begin{subequations}\label{TPV}
\bea
|\Ga\Lb \nu,i\nu_1=\h,i\nu_2=\h\Rb|^2&&=\h\Lb 4\pi\Rb^6\bas^4\Lb 1-\f{1}{N_c^2}\Rb^2\Lb\f{1}{4}+\nu^2\Rb^3\chi\Lb\nu\Rb\f{1-i\nu_1-i\nu_2}{\Lb 1/2-i\nu_1\Rb^2\Lb 1/2-i\nu_2\Rb^2}\hspace{0.5cm}\label{TPV1}\\
|\Ga\Lb i\nu=\h,\nu_1=0,\nu_2=0\Rb|^2&&=\f{\Lb 4\pi\Rb^6\bas^4}{N_c^4}\f{1}{\Lb 1/4+\nu^2\Rb}\label{TPV2}\eea
\end{subequations}

\eq{TPV1} is dominated by the planar diagram which contains singularities in region I, and the non planar diagram which
is non singular has been thrown away. \eq{TPV2} is purely the contribution from the non planar diagram which contains singularities in region II, whereas the planar
diagram is non singular in region II and is not included in \eq{TPV2}. Previous publications assumed that $N_c\to\infty$, so since the non planar diagram has a pre-factor of $1/N_c^2$,
it was neglected. In this letter the non planar diagram has been taken into account and moreover it is the dominant contribution to the TPV for region II. The non planar TPV 
leads to the contribution to the loop amplitude of \fig{fsopl} (a) derived in \eq{A2} below, which is the dominant contribution. Hence in this letter, the remarkable property of the Pomeron loop amplitude 
has been found, that the major contribution to the Pomeron loop amplitude stems from the non planar TPV. 

\eq{TPV1} and \eq{TPV2} lead to two contributions to the Pomeron loop amplitude for regions I and II. For region I the $\nu_1,\nu_2$ integrals
 in \eq{loop2} are evaluated by closing the contour on the upper half plane and summing over the residues at $i\nu_1,i\nu_2=1/2$, and for region II the steepest descent method is used (see ref.  \cite{Miller:2009ca} where
 the details of this calculation are explained).
 The loop amplitude $d_{(1)}\Lb\nu|y\Rb$ is the sum of both contributions given in \eq{third} and \eq{fourth}, namely  \cite{Miller:2009ca};

\begin{subequations}\label{grp} 
\bea
d_{(1)}\Lb\nu|y,\de y_H\Rb&&=d\Lb\f{1}{4}+\nu^2\Rb^3\chi\Lb\nu\Rb\,e^{-\omega\Lb\nu\Rb\de y_H }\Lb\omega\Lb\nu\Rb+\h\omega^2\Lb\nu\Rb \Lb y-\de y_H\Rb\Rb\label{second}\\ 
&&+\f{a e^{-\omega\Lb 0\Rb \de y_H}}{\Lb 1/4+\nu^2\Rb }\int^y_{\de y_H}\!\!\! dy_1\!\int^{y-\de y_H}_0\!\!\!\!\!\! dy_2 \f{e^{\Lb 2\omega\Lb 0\Rb-\omega\Lb\nu\Rb\Rb y_{12}}}{y_{12}^{3/2}\Lb y_{12}-\de y_H\Rb^{3/2}};\label{third}\\
d&&=\bas\Lb 1-\f{1}{N_c^2}\Rb^2;\hspace{0.5cm}a=\f{2^{10}\bas^4}{N_c^4\pi\Lb\omega^{\prime\prime}\Lb 0\Rb\Rb^3}\,.\label{fourth}
\eea
\end{subequations} 

Plugging \eq{grp} into \eq{loop1} leads to two contributions for the scattering amplitude of \fig{fsopl} (a). The first part stems from \eq{second} for region I, and the $\nu$ integral is evaluated by the
method of steepest descents. The second piece stems from \eq{third} for region II, for which the $\nu$ integral is solved by closing the contour over the upper
half plane and summing over the residues at $i\nu=1/2$, and afterwards integrating over the rapidity variables to yield finally the two expressions  \cite{Miller:2009ca};

\begin{subequations}\label{grp1} 
\bea
A_{(1)}\Lb y,\de y_H|\mbox{\fig{fsopl} (a)}\Rb&&=\f{\Lb 2\pi\Rb^{1/2}\bas^2 d A_H}{128\pi^2 N_c^2}\f{e^{\omega\Lb 0\Rb\Lb y-\de y_H\Rb}}{\Lb\omega^{\prime\prime}\Lb 0\Rb\Lb y-\de y_H\Rb\Rb^{3/2}}\chi\Lb 0\Rb
\Lb\omega\Lb 0\Rb+\h\omega^2\Lb 0\Rb\Lb y-\de y_H\Rb\Rb\hspace{0.5cm}\label{A1}\\
&&+\f{\bas a A_H}{2^9\pi N_c^2}\Lb y-\de y_H\Rb\Lb\f{-1}{\bas}\f{d}{dy}\Rb^3\f{e^{2\omega\Lb 0\Rb\Lb y-\de y_H/2\Rb}}{y^{3/2}\Lb y-\de y_H\Rb^{3/2}}\label{A2}
\eea
\end{subequations} 

\eq{A1} is the part of the scattering amplitude which leads to the renormalization of the Pomeron intercept $\omega\Lb 0\Rb$. \eq{A2} is the dominant contribution,  and is equivalent to 2 non interacting Pomerons, with renormalized Pomeron vertices. This can be seen from \fig{fsopl} (a), whereby at high energy taking out both branches of the loop, one is left with just two non interacting Pomeron exchanges. The same is true of the multiple loop diagrams
of \fig{fsopl} (b) and (c), and all higher order symmetric loop diagrams, as will now be shown. \\

As explained above, \fig{fsopl} (b) stems from \fig{fsopl} (a) when each branch of the loop gives birth to a secondary loop leading to the two ``second generation" of loops in  \fig{fsopl} (b).
 In the same way when the second generation loops in \fig{fsopl} (b) each give birth to two loops, this leads to 4 ``third generation" of loops in \fig{fsopl} (c). Continuing with this evolution, the entire spectrum of symmetric $N$ generation diagrams can be generated for all $N$ . The scattering amplitude with $N$ generations of loops shown in \fig{fpictN1}
is the generalization of \eq{loop1}, namely  \cite{Miller:2009ca};

\bea
A_{(N)}\Lb y,\de y_H|\mbox{\fig{fpictN1}}\Rb&&=\f{\as^2}{4}\If\!\!\! d\nu h\Lb\nu\Rb \la^2\Lb\nu\Rb e^{\omega\Lb\nu\Rb y}d_{(N)}\Lb\nu|y,\de y_H\Rb E_\nu E^\prime_{-\nu} A_H\label{AN}\eea

\FIGURE[h]{\begin{minipage}{150mm}
\centerline{\epsfig{file=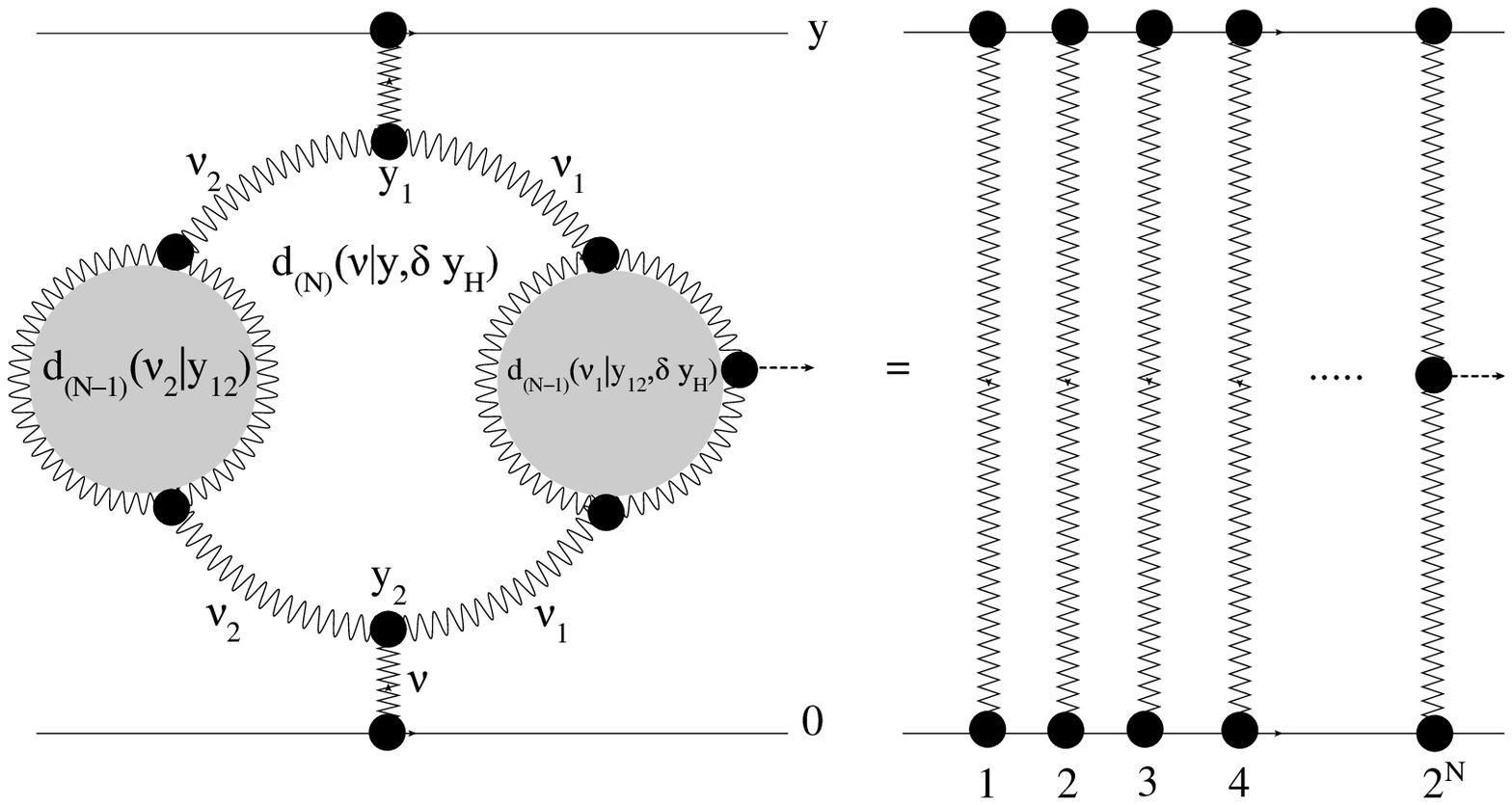,width=150mm}}
\end{minipage}
\caption{The $N$ generation diagram which stems from the simple loop giving
birth to two sets of $N-1$ generations of loops, is equivalent to the diagram of $2^N$ non interacting Pomerons with renormalized Pomeron vertices.} \label{fpictN1} } 

where $d_{(N)}\Lb\nu|y,\de y_H\Rb$ is the contribution of the $N$ generations of loops in \fig{fpictN1}. The way to calculate $d_{(N)}\Lb\nu|y,\de y_H\Rb$
is based on the observation that \fig{fpictN1} is equivalent to the simple loop diagram of \fig{fsopl} (a) when each branch in the loop gives birth to a set of $N-1$ generations of loops. This means
that to write the expression for $d_{(N)}\Lb\nu|y,\de y_H\Rb$, all that is needed is to modify the propagators for the branches of the loop in the expression of \eq{loop2} as;

\vspace{1cm}

\bea
\la\Lb\nu_1\Rb\to\la\Lb\nu_1\Rb d_{(N-1)}\Lb\nu_1|y_{12},\de y_H\Rb\la\Lb\nu_1\Rb;\hspace{0.5cm}
\la\Lb\nu_2\Rb\to\la\Lb\nu_2\Rb d_{(N-1)}\Lb\nu_2|y_{12}\Rb\la\Lb\nu_2\Rb\label{lamodified}\eea

where in the notation of \fig{fpictN1}, $d_{(N-1)}\Lb\nu_1|y_{12},\de y_H\Rb$ labels the set of $N-1$ generations of loops on the right in \fig{fpictN1} which includes the
production of the Higgs, and $d_{(N-1)}\Lb\nu_2|y_{12}\Rb=d_{(N-1)}\Lb\nu_2|y_{12},\de y_H=0\Rb$ labels the set of loops on the left in \fig{fpictN1} where there is no Higgs production.
After implementing \eq{lamodified} in \eq{loop2}, one arrives at the following  amplitude for the set of $N$ generations of loops  \cite{Miller:2009ca};

\bea
d_{(N)}\Lb\nu|y,\de y_H\Rb&&=\f{1}{16}\If\!\!\! d\nu_1 h\Lb\nu_1\Rb\la\Lb\nu_1\Rb\If\!\!\! d\nu_2 h\Lb\nu_2\Rb\la\Lb\nu_2\Rb|\Ga\Lb\nu|\nu_1,\nu_2\Rb|^2\label{loopN}\\
&&\times\int^y_{\de y_H}\!\!\! dy_1\!\int^{y-\de y_H}_0\!\!\!\!\!\! dy_2 e^{\Lb\omega\Lb\nu_1\Rb+\omega\Lb\nu_2\Rb-\omega\Lb\nu\Rb\Rb y_{12}-\omega\Lb\nu_1\Rb\de y_H}
d_{(N-1)}\Lb\nu_1|y_{12},\de y_H\Rb d_{(N-1)}\Lb\nu_2|y_{12}\Rb\nn
 \eea

\eq{loopN} forms an iterative expression. Using the technique of proof by induction, the following two formulae for the two contributions to \eq{loopN} can be proved (see ref.  \cite{Miller:2009ca}) which stem from 
the two asymptotic expressions for the TPV of \eq{TPV1} and \eq{TPV2};

\begin{subequations}\label{dN1}
\bea
&&d^{\,\footnotesize{\mbox{I}}}_{(N)}\Lb \nu|y,\de y_H\Rb\,
=\f{\Lb d\,d^{\,\prime}\Rb^{2^{[N-1]}}\bas^{N-1}}{d^{\,\prime}}\Lb\f{1}{4}+\nu^2\Rb^3\chi\Lb\nu\Rb\,e^{-\omega\Lb\nu\Rb \de y_H}\label{dN2}\\
&&\hspace{2.6cm}\times
\prod^{N}_{k=2}\bas^{\Lb k-1\Rb2^{[N-k]}}
\Lb\f{k+1}{k}-\f{\Lb k+3\Rb\Lb k+2\Rb\Lb k+1\Rb}{2 k^2}\Rb^{2^{[N-k]}-1}\,\nn\\
&&\hspace{2.6cm} \times\prod^{N}_{k=2}\Lb\f{k+1}{k}-\f{\Lb k+3\Rb\Lb k+2\Rb\Lb k+1\Rb}{2 k^2}+\f{k+1}{k}\de y_H\Rb\,\nn\\
&&\hspace{2.6cm}\times\,\,\omega^{N-1}\Lb \nu\Rb\Lb\omega\Lb \nu\Rb+\f{\omega^2\Lb \nu\Rb\Lb y_1-\de y_H\Rb}{N+1}\Rb\hspace{5cm}\forall N\geq1;\nn\\
\nn\\
&&d^{\mbox{\footnotesize{II}}}_{(N)}\Lb\nu|y,\de y_H\Rb=
\f{\Lb ab\Rb^{2^{[N-1]}}}{b}\Lb \f{1}{4}+\nu^2\Rb^3\chi\Lb\nu\Rb e^{-\omega\Lb \nu\Rb\de y_H}\int^y_{\de y_H}\!\!\!\!\! dy_1\int^{y_1-\de y_H}_{0}\!\!\!\!\! \!\!dy_2\, y_{12}^{2^{[N-1]}-N}\Lb y_{12}-\de y_H\Rb^{N-1}\hspace{0.8cm}\label{dN3}\\
&&\hspace{2.6cm}\times
\left\{\!\Lb\f{-1}{\bas}\f{d}{dy_{12}}\Rb^3\!\f{e^{2\omega\Lb 0\Rb\Lb y_{12}-\de y_H/2\Rb}}{y_{12}^{3/2}\Lb y_{12}-\de y_H\Rb^{3/2}}\!\right\}\!
\left\{\!\Lb\f{-1}{\bas}\f{d}{dy_{12}}\Rb^3\!\f{e^{2\omega\Lb 0\Rb y_{12}}}{y_{12}^3}\!\right\}^{2^{[N-1]}-1}\hspace{0.5cm}\forall N\geq 1;\nn\\
\nn\\
&&d=\bas\Lb 1-\f{1}{N_c^2}\Rb^2;\hspace{0.5cm}d^{\,\prime}=\f{\bas}{8}\Lb 1-\f{1}{N_c^2}\Rb^2;\hspace{0.5cm}
a=\f{2^{10}\bas^4}{N_c^4\pi[\omega^{\,\prime\prime}\Lb 0\Rb]^3};\hspace{0.5cm}
b=\f{\bas^2}{2^{11}}\Lb 1-\f{1}{N_c^2}\Rb^2.\label{fullsetofconstants}
\eea
\end{subequations}

Finally after inserting \eq{dN1} into \eq{AN}, one finds the following scattering amplitude for the diagram of \fig{fpictN1} with $N$ generations of loops  \cite{Miller:2009ca};

\begin{subequations}\label{AN1}
\bea
A^{\,\footnotesize{\mbox{I}}}_{(N)}\Lb y,\de y_H|\mbox{\fig{fpictN1}}\Rb\,
&&=\f{\Lb 2\pi\Rb^{1/2}\bas^2A_H}{128\pi^2N_c^2}\f{\Lb d\,d^{\,\prime}\Rb^{2^{[N-1]}}}{d^{\,\prime}}\bas^{N-1}\chi\Lb0\Rb
\f{e^{\omega\Lb 0\Rb\Lb y- \de y_H\Rb}}{\Lb\omega\Lb 0\Rb^{\prime\prime}\Lb y-\de y_H\Rb\Rb^{3/2}}\label{AN2}\\
&&\times
\prod^{N}_{k=2}\bas^{\Lb k-1\Rb2^{[N-k]}}
\Lb\f{k+1}{k}-\f{\Lb k+3\Rb\Lb k+2\Rb\Lb k+1\Rb}{2 k^2}\Rb^{2^{[N-k]}-1}\,\nn\\
&& \times\prod^{N}_{k=2}\Lb\f{k+1}{k}-\f{\Lb k+3\Rb\Lb k+2\Rb\Lb k+1\Rb}{2 k^2}+\f{k+1}{k}\de y_H\Rb\,\nn\\
&&\times\,\,\omega^{N-1}\Lb 0\Rb\Lb\omega\Lb 0\Rb+\f{\omega^2\Lb 0\Rb\Lb y_1-\de y_H\Rb}{N+1}\Rb\,\hspace{3.5cm}\forall N\geq 1;\hspace{0.5cm}\nn\\
\nn\\
\nn\\
A^{\mbox{\footnotesize{II}}}_{(N)}\Lb y,\de y_H|\mbox{\fig{fpictN1}}\Rb&&=\f{\bas\,A_H}{2^9 N_c^2\pi}
\f{\Lb ab\Rb^{2^{[N-1]}}}{b}y^{2^{[N-1]}-N}\Lb y-\de y_H\Rb^N\label{AN3}\\
&&\times 
\left\{\Lb\f{-1}{\bas}\f{d}{dy}\Rb^3\f{e^{2\omega\Lb 0\Rb \Lb y-\de y_H/2\Rb}}{y^{3/2}\Lb y-\de y_H\Rb^{3/2}}\!\right\}\!
\left\{\!\!\Lb\f{-1}{\bas}\f{d}{dy}\Rb^3\!\!\f{e^{2\omega\Lb 0\Rb y}}{y^3}\!\!\right\}^{2^{[N-1]}-1}\forall N\geq 1.\hspace{1.2cm}\nn\eea
\end{subequations}

\eq{AN2} is the contribution which leads to the renormalization of the Pomeron intercept. \eq{AN3} is the piece which is equivalent to $2^N$ non interacting Pomerons, with renormalized Pomeron vertices
shown pictorially in \fig{fpictN1}. This follows from the observation that \eq{AN3} can be recast in the form;

\bea
A^{\mbox{\footnotesize{II}}}_{(N)}\Lb y,\de y_H|\mbox{\fig{fpictN1}}\Rb\equiv \ka_{\mbox{\tiny{\it (N)}}}\,e^{2^N\omega\Lb 0\Rb y}\label{noninteratingpomerons}\eea

Comparison of the RHS of  \eq{noninteratingpomerons}with \eq{A01} which is proportional to $e^{\omega\Lb 0\Rb y}$, shows that \\
$A^{\mbox{\footnotesize{II}}}_{(N)}\Lb y,\de y_H|\mbox{\fig{fpictN1}}\Rb$ is equivalent to the amplitude of $2^N$ non interacting Pomerons, with a pre-factor $\ka_{\mbox{\tiny{\it (N)}}}$
which contains the set of renormalized Pomeron vertices. 
The complete scattering amplitude is the sum of \eq{AN2} and \eq{AN3}.

\TABLE[ht]{
\begin{tabular}{||c|c||}
\hline \hline
&\\ 
   & $A_{(N)}\Lb y=19,\de y_H=\ln\Lb M_H^2/4 s_0\Rb \Rb$ \\
 & $\as=0.12$\hspace{1.8cm}$\as=0.2$\\
 \hline
$N=0$ & $1.57\times10^{-8}\,\,\,\,$$\hspace{1cm}\,\,2.17\times10^{-7}$ \\ \hline
$N=1$&   $2.05\times10^{-10}\,\,$$\hspace{1cm}\,\,2.15\times10^{-7}$\\ \hline
$N=2$   
& $2.85\times10^{-15}\,\,$$\hspace{1cm}\,\,1.39\times10^{-7}$ \\
\hline
$N=3$  
 & $3.40\times10^{-24}\,\,$$\hspace{1cm}\,\,1.28\times10^{-7}$\\
 \hline
$N=4$  
& $8.20\times10^{-42}\,\,$$\hspace{1cm}\,\,2.31\times10^{-8}$\\ \hline
$N=5$  
 & $8.13\times10^{-77}\,\,$$\hspace{1cm}\,\,5.11\times10^{-9}$ \\
\hline
$N=6$ 
 & $1.36\times10^{-146}\,\,$$\hspace{1cm}4.26\times10^{-10}$   \\ 
\hline \hline
\end{tabular}
  \caption{
Scattering amplitudes  \cite{Miller:2009ca} for diffractive Higgs
production from the multiple Pomeron loop diagram with $N$ generations of loops. The mass of the Higgs boson
is assumed to be $M_H=100\,\mbox{GeV}$, and the rapidity gap $y$ between the scattering protons is taken to be $y=19$, based on
proton proton collisions at the typical LHC energy $\sqrt{s}=14$ TeV.
 }\label{t3}}

\Tab{t3}  \cite{Miller:2009ca} shows the results up to the $N=6$ generation diagram. $N=0$ refers to the basic amplitude of \fig{flrg1} without loops derived in
\eq{A01}. The values in the table show that
for the range of energies relevant to the LHC, the scattering amplitude falls as the number of generations of loops introduced increases. 
However from $N=0$ to $N=4$, for $\as=0.2$ the scattering amplitudes are the same order of magnitude. This demonstrates the importance
of taking into account the full set of Pomeron loops for the summation over Pomeron loop diagrams.
The difference in values for different choices of $\as$ show that the scattering amplitude is very sensitive to the choice of intercept. \\

\vspace{1.5cm}

The survival probability is a quantitative measure of the shadowing corrections to the basic diagram of \fig{flrg1}.  This stems from additional parton showers 
which produce inelastic scattering  that destroy the large rapidity gaps. The contribution of hard re-scattrering to the survival probability can be derived
from summing over the additional parton showers that stem from the Pomeron loop diagrams which have been calculated above. In this way, the 
``enhanced survival probability" $\SPE$ is calculated by subtracting from the basic diagram of \fig{flrg1}, the first correction from the $N=1$ generation diagram of
\fig{fsopl} (a), and subtracting from this the $N=2$ generation diagram of \fig{fsopl} (b) , and so on, and divide by the basic amplitude of \fig{flrg1} itself to yield the
correctly normalized survival probability which leads to the formula  \cite{Miller:2009ca};

\bea
\SPE=1-\sum^\infty_{N=1}\Lb -1\Rb^{N-1}\f{A_{(N)}\Lb y,\de y_H|\mbox{\eq{AN2}+\eq{AN3}}\Rb}{A_{(0)}\Lb y,\de y_H|\mbox{\eq{A01}}\Rb}\eea

.  \FIGURE[h]{\begin{minipage}{80mm}
\centerline{\epsfig{file=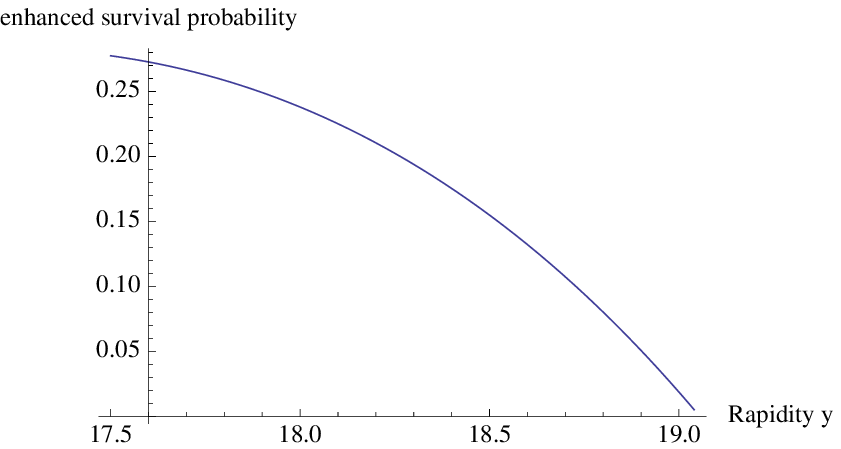,width=80mm}}
\end{minipage}
\caption{The enhanced survival probability $\SPE$  plotted against the rapidity separation $y$ of the scattering protons. $\as = 0.2$. } \label{fspgraph} } 
\fig{fspgraph}   \cite{Miller:2009ca} is a plot of the energy dependence of the enhanced survival probability. The graph shows a decrease of the survival probability as the energy increases. For
the LHC range of energies, the results show that the enhanced survival probability is small, and could even be less than 1\%. Both of these observations are in agreement with
the results found in refs. \cite{Miller:2006bi,Gotsman:2009vz,Gotsman:2009bn,Gotsman:2008tr}. The decrease with energy is a natural consequence of the way the survival probability is defined.
The survival probability measures the probability not to have  extra parton showers other than the basic process of \fig{flrg1}. Extra parton showers leading to 
unwanted inelastic scattering  inevitably increase with the
energy of the interaction, and this decreases the probability of exclusive Higgs production with large rapidity gaps in tact after the scattering. \\

In summary, the dominant contribution to the simple Pomeron loop of \fig{fsopl} (a) stems from the non planar triple Pomeron vertex shown in 
\fig{ftpv} (b), which has not been taken into account before. The formula for the multiple  Pomeron loop diagram
for  $N$ generations of loops, has been derived in the perturbative QCD approach, whereas previous methods have relied on
the mean field approximation. The multiple Pomeron loop amplitude has two contributions which lead to the renormalization
of the Pomeron intercept, and the diagram of $2^N$ non interacting Pomerons for the $N$ generation diagram, with renormalized Pomeron vertices. 
Multiple Pomeron loop diagrams, other than the simple loop of \fig{fsopl} (a) contribute significantly to the summation over Pomeron loops, and need to be taken into account.
Using this formula, the contribution of short distance interactions to the survival probability has been estimated,
from the sum over Pomeron loop diagrams, in perturbative QCD. The survival probability decreases with energy, and for the LHC range of energies it 
is small and could even be less than 1\%., in agreement with the findings of refs. \cite{Miller:2006bi,Gotsman:2009vz,Gotsman:2009bn,Gotsman:2008tr}.
The smallness of the enhanced survival probability show that short distance interactions contribute substantially.\\

We would like to
thank E. Levin and G. Milhano for their careful reading and helpful advice in writing this paper.  I would
also like to thank L. Apolin$\acute{a}$rio, M. Braun, J. Dias De Deus, and G. Milhano  for
fruitful discussions on the subject. 
This research was supported by the Funda\c{c}$\tilde{a}$o para ci$\acute{e}$ncia e a tecnologia (FCT), and CENTRA - Instituto Superior T$\acute{e}$cnico (IST), Lisbon.

\end{document}